\documentclass[a4paper,12pt,dvips]{article}

\usepackage{t1enc}
\usepackage[latin1]{inputenc}

\usepackage{amsmath,amssymb,amsfonts,amsthm} 
\usepackage{graphicx}
\newtheorem{theorem}{Theorem}[section]

\newcommand{\spa}[1]{\mathcal{#1}}

\newcommand{\lt}{L^{2}(\Omega)}

\newcommand{\hk}[1]{H^{#1}(\Omega)}

\newcommand{\hkl}[1]{H^{#1}_{loc}(\Omega)}

\newcommand{\hfk}[2]{||#1||_{H^{#2}(\Omega)}}

\newcommand{\Rt}{\mathbb{R}^3}

\newcommand{\ck}{\mathcal{L}_h}

\newcommand{\Lh}{\mathbf{L}_h}

\newcommand{\Oi}{\Omega\setminus \{i\}}

\newcommand{\B}{\mathbf{B}}

\newcommand{\Qs}[1]{Q^{#1}_{sing}}
\newcommand{\Qr}[1]{Q^{#1}_{reg}}

\title{Trapped surfaces as boundaries for the constraint equations}
\author{Sergio Dain\\
  Max-Planck-Institut f\"ur Gravitationsphysik\\
  Am M\"uhlenberg 1\\
  14476 Golm\\
  Germany}

\begin{document}
\maketitle

\begin{abstract}
  Trapped surfaces are studied as inner boundary for the Einstein
  vacuum constraint equations. The trapped surface condition can be
  written as a non linear boundary condition for these equations.
  Under appropriate assumptions, we prove existence and uniqueness of
  solutions in the exterior region for this boundary value problem.
  We also discuss the relevance of this result for the study of  black holes
  collisions.

\end{abstract}

\section{Introduction}
A black hole defines a boundary in the spacetime: the boundary of the
region of no escape to infinity. This boundary is called the event
horizon.  There exist others kinds of
spacetime boundaries, for example the ones defined by matter sources:
 the interface between matter and vacuum. In this case the
boundary is introduced in the sources of Einstein equations by choosing
an energy density with compact support. For black holes, the boundary
is produced by the vacuum equations themselves, it depends only on
fundamental properties of gravity; in this sense it is  a  more
fundamental kind of boundary than the matter sources ones.

In the presence of a boundary, is natural to study boundary conditions
for the equations and solve only for the exterior region.  In the case
of black holes this is very desirable since, firstly, the observations
are made at infinity which is causally disconnected with the black hole
region. And, secondly, in the black hole region there are singularities
which are very difficult to handle in the numeric simulations.

In the context of an initial value formulation, the first step in
order to understand the black hole boundary value problem is the study
of the intersection of this boundary with a spacelike
three-dimensional Cauchy hypersurface. That is, to study the black
hole boundary value problem for the constraint equations. This will be
the subject of this article.

There is no known way to characterize the intersection of the event
horizon with a Cauchy surface in terms of a differential condition 
which involves only fields on the Cauchy surface.  Such local
characterization is provided by the concept of trapped surface.  A
trapped surface is a compact, two-dimensional surface for which the
expansions of both sets, outgoing and ingoing, of future directed
null geodesics orthogonal to the surface are negative. The relevance
of trapped surfaces relies in two important results: i) a trapped
surface is always inside an event horizon. ii) The development of an
initial data set which contain a trapped surface will be geodesically
incomplete (see \cite{Hawking73} and \cite{Wald84}).  Because of i) we
expect that trapped surfaces should be good boundaries for the
constraint equations. In this article we will see that in fact this is
true: trapped surfaces provide inner boundary conditions for
the constraint equations under appropriate assumptions.

Associated with trapped surfaces there is the concept of an apparent
horizon.  An apparent horizon is essentially defined as the outermost
trapped surface. In this article only trapped surfaces will be
discussed and not apparent horizons, since with the techniques used
here it will allows to decide whether a given trapped surface is in
fact the outermost.

The problem of constructing initial data with several disconnected
trapped surfaces is relevant for the study of black holes collisions
(see for example \cite{Cook00} and \cite{Dain02g}).  In almost all
data currently used in numerical simulations (the exceptions are the
numerical studies in \cite{Thornburg87} and \cite{Husa96}) trapped
surfaces are produced by indirect means, for example introducing a non
trivial topology. The theorem proved in this article will provide a
way of solving only for the exterior region, with the appropriate
boundary conditions, without imposing any symmetry and without
introducing any non trivial topology.

From the point of view of getting initial data for black hole
collisions this has several advantages.  Firstly, no computer
resources are wasted in the interior region.  Secondly, the location
of the trapped surfaces is known a priori.  Finally, the control of
the boundary conditions presented here will allow one to construct more
general classes of black hole initial data than the ones studied so
far (see the discussion in section \ref{main}).

The plan of the article is the following. In section \ref{main} we
present our result, given by theorems \ref{T1} and \ref{T2}. We also
discuss the physical interpretation of these theorems. In section
\ref{pre} we collect some results from the elliptic theory that will
be used in the proofs. In section \ref{mc} and section \ref{hc} we
prove theorem \ref{T1} and theorem \ref{T2} respectively.

\section{Main Result}\label{main}

Let $C_k$ be a finite collection of \emph{compact} sets in $\mathbb{R}^3$. We
define the exterior region $\tilde \Omega=\mathbb{R}^3\setminus \cup_k
C_k $. An \emph{initial data set} for the Einstein \emph{vacuum} equations is
given by the triple $(\tilde \Omega, \tilde h_{ab}, \tilde K_{ab})$
where $\tilde h_{ab} $ is a (positive definite) Riemannian metric, and
$\tilde K_{ab}$ a symmetric tensor field on $\tilde \Omega$, such that
they satisfy the constraint equations
\begin{equation}
 \label{const1}
\tilde D^b \tilde K_{ab} -\tilde D_a \tilde K=0,
\end{equation}
\begin{equation}
 \label{const2}
\tilde R + \tilde K^2-\tilde K_{ab} \tilde K^{ab}=0,
\end{equation}
on $\tilde \Omega$; where $\tilde D_a$ is the covariant derivative
with respect to $\tilde h_{ab}$, $\tilde R$ is the trace of the
corresponding Ricci tensor, and $\tilde K=\tilde h^{ab} \tilde
K_{ab}$.

The data will be called \emph{asymptotically flat} if there exists some
 compact set $C$, with  $\cup_k C_k \subset C$, such that   $\tilde
\Omega \setminus C$ can be mapped by a coordinate system $\tilde x^j$
diffeomorphically onto the complement of a closed ball in
$\mathbb{R}^3$ and we have in these coordinates
\begin{equation} 
\label{pf1}
\tilde h_{ij}=(1+\frac{2m}{\tilde r})\delta_{ij}+O(\tilde r^{-2}),
\end{equation}
\begin{equation} 
\label{pf2}
\tilde K_{ij}=O(\tilde r^{-2}),
\end{equation}
as $\tilde r= ( \sum_{j=1}^3 ({\tilde x^j})^2 ) ^{1/2} \to \infty$,
where the constant $m$ is the total mass of the initial data.

\begin{figure}[htbp]
  \begin{center}
    \includegraphics[width=8cm]{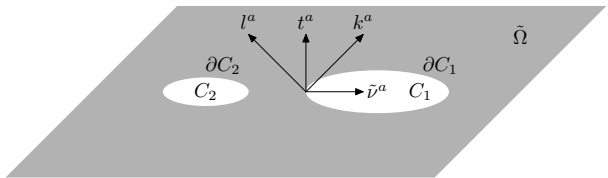}
    \caption{The exterior region $\tilde\Omega$  with two boundary
      components $\partial C_1$ and $\partial C_2$.}
    \label{fig:1}
  \end{center}
\end{figure}

The boundaries $\partial C_k$ are assumed to be smooth, two
dimensional surfaces in $(\tilde \Omega, \tilde h )$. Let $\tilde
\nu^a$ be the unit normal of $\partial C_k$, with respect to $\tilde
h_{ab}$, pointing in the \emph{outward} direction of $\tilde \Omega$.
Let $t^a$ be the unit timelike vector field orthogonal to the
hypersurface $\tilde \Omega$ with respect to the spacetime metric
$g_{ab}$ ($t^at^b g_{ab}=-1$ with our signature convention) The
outgoing and ingoing null geodesics orthogonal to $\partial C_k$ are
given by $ l^a=t^a-\tilde \nu^a $ and $k^a=t^a+\tilde \nu^a$
respectively, the corresponding expansions are given by
$\Theta_+=\nabla_a l^a$ and $\Theta_-=\nabla_a k^a$, where $\nabla_a$
is the connexion with respect to $g_{ab}$, see Fig.  \ref{fig:1}.  We
can calculate these expansions in terms of quantities intrinsic to the
initial data
\begin{align}
  \label{eq:21}
  \Theta_- &=\tilde K +\tilde H -\tilde \nu^a \tilde
  \nu^b \tilde K_{ab},\\
\Theta_+ &=\tilde K - \tilde H-\tilde \nu^a \tilde
  \nu^b \tilde K_{ab}.\label{eq:21b}
\end{align}
where $\tilde H = \tilde D_a\tilde \nu^a$.  Note that $\tilde
K=\nabla_a t^a$ is the mean curvature of the three-dimensional
hypersurface $\tilde \Omega$ with respect to the spacetime metric
$g_{ab}$ and the normal $t^a$, and $\tilde H$ is the mean curvature of
the two-dimensional surface $\partial C_k$ with respect to
the metric $\tilde h_{ab}$ and the normal $\tilde \nu^a$. 
Our choice of the definition of the extrinsic curvature is given by
$\tilde K_{ab}=\tilde h^c_a \nabla_c t_b$ which agree with
\cite{Wald84} but is the negative of the choice made in
\cite{Misner73}.  If $\partial C$ is an sphere of radius $\tilde r$ on
a canonical Minkowski slice $t=const.$, then we have $\Theta_+=2/\tilde r$,
$\Theta_-=-2/\tilde r$, $\tilde H = -2/\tilde r$, $\tilde K_{ab}=0$,
and $\tilde v^a=-(\partial/\partial \tilde r)^ a$.

The boundary $\partial C_k$ will be called a \emph{future trapped
  surface} if $\Theta_+ <0$ and $\Theta_- <0$ on $\partial C_k$ and a
\emph{future marginally trapped surface} if $\Theta_+ \leq 0$ and
$\Theta_- \leq 0$.

We want to find solutions of the constraint equations
(\ref{const1})--(\ref{const2}) which are asymptotically flat (i.e;
they satisfy (\ref{pf1})--(\ref{pf2})) and such that all the boundary
components $\partial C_k$ (i.e; the whole boundary $\partial \tilde
\Omega$) are trapped surfaces. In order to achieve this, we will
reduce the constraint equations to an elliptic boundary value problem
in which a negative (non-positive) ingoing null expansion $\Theta_-$
can, essentially, be freely prescribed at the boundary $\partial
\tilde \Omega$. Under further restrictions on $\Theta_-$ it will
follows that also $\Theta_+$ will be negative (non-positive), and
hence the boundary will be future trapped (future marginally trapped).
We emphasize that only $\Theta_-$ (and not $\Theta_+$) will be 
free data on the boundary.
 
The elliptic reduction will be given by the conformal method.  Also,
instead of working with the exterior region $\tilde\Omega$ it will be
more convenient to work with its related compactification $\Omega$. In
the following we describe both procedures (see \cite{Choquet99},
\cite{Choquet80} and the references therein for a description of the
conformal method, and references \cite{Beig94}, \cite{Friedrich88},
\cite{Friedrich98}, \cite{Dain99} for the compactification
procedure).

\begin{figure}[htbp]
  \begin{center}
    \includegraphics[width=5cm]{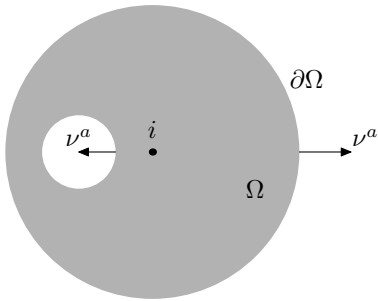}
    \caption{The compactification  $\Omega$ of the exterior region
      $\tilde \Omega$ showed in Fig. \ref{fig:1}.}
    \label{fig:2}
  \end{center}
\end{figure}

Let $\Omega$ a bounded domain in $\Rt$. We will assume that $\Omega$
is connected, however in general it will not be simply connected,
each hole in $\Omega$ will correspond to an extra boundary component
$\partial C_k$. Let $i\in \Omega$ be an arbitrary point and set
$\tilde \Omega= \Oi$. The point $i$ will represent the infinity of
$\tilde \Omega$.  Let $h_{ab}$ be a Riemannian metric defined on $\bar
\Omega$. We denote by $\nu^a$ the unit normal to $\partial \Omega$
with respect to $h_{ab}$, pointing in the \emph{outward} direction, see Fig.
\ref{fig:2}.  We will assume that the conformal metric satisfies
\begin{equation}
  \label{eq:h-reg}
  h_{ab}\in C^\infty(\bar \Omega \setminus \{i \} )\cap
  C^2(\bar \Omega).
\end{equation}
We make different assumptions concerning the regularity of $h_{ab}$ at
$\bar \Omega \setminus \{i \}$ and at $i$.  The regularity assumption
at $i$ is concerned with the fall off behavior of the solution at
infinity, it is not related with the inner boundaries $\partial C_k$.
The physical motivation for this distinction has been discussed in
\cite{Dain02b}, \cite{Dain02d} and \cite{Dain01b}; in particular
assumption \eqref{eq:h-reg} is weaker that the ones made in these
references.

Let $K^{ab}$ be a symmetric, trace free tensor with respect to $h_{ab}$,
such that
\begin{equation}  
\label{diver}
D_a K^{ab}=0 \text{ on } \tilde \Omega,
\end{equation}
and let $\psi$ be a positive solution of 
\begin{equation} 
\label{Lich}
L_h \psi=-\frac{1}{8}K_{ab}K^{ab}\psi^{-7} \text{ on } \tilde \Omega, 
\end{equation}
where $L_h\equiv D^aD_a-R/8$ and $R$ is the Ricci scalar of the metric
$h_{ab}$.  If we are able to find solutions of equations
\eqref{diver}--\eqref{Lich}, then the physical fields $(\tilde h,
\tilde K)$ defined by $\tilde{h}_{ab} = \psi^4 h_{ab}$ and
$\tilde{K}^{ab} = \psi^{-10}K^{ab}$ will satisfy equations
\eqref{const1}--\eqref{const2} on $\tilde \Omega$. Since we have
assumed $K=0$ we will obtain $\tilde K=0$.

We discuss now boundary conditions for equations
\eqref{diver}--\eqref{Lich}. There are two kind of boundary
conditions. The first one is asymptotic flatness,  we want to
ensure that the solution will satisfies the fall off
\eqref{pf1}--\eqref{pf2}. For this we require
\begin{equation}
  \label{eq:77}
  K^{ab}=O(r^{-4}) \text{ at } i,
\end{equation}
and
\begin{equation}
  \label{eq:17}
\lim_{r\to 0} r\psi=1   \text{ at } i,
\end{equation}
where $x^i$ are normal coordinates with respect to the metric $h_{ab}$ 
centered at $i$ and $r$ the corresponding radius. 

The second boundary condition is the requirement that $\Theta_-$
should be freely prescribed at the boundary $\partial \Omega$. In order to
write this condition we first calculate the null expansions $\Theta_+$
and $\Theta_-$  in term of the conformal quantities
\begin{align}
\label{eq:t+}
  \Theta_- &= \psi^{-3}\left(4 \nu^aD_a
    \psi + H \psi  - \psi^{-3} K_{ab} \nu^a \nu^b \right),\\
\label{eq:t-}
\Theta_+ &= \psi^{-3}\left(  -4 \nu^aD_a
    \psi -  H\psi  - \psi^{-3} K_{ab} \nu^a \nu^b \right),
\end{align}
where we have used that $K=\tilde K=0$, the normal vectors are related
by $\tilde \nu^a=\psi^{-2}\nu^a$ (that is, $\nu^a$ is a unit vector
with respect to the conformal metric $h_{ab}$) and $H=D_a\nu^a$. From equation
\eqref{eq:t+} we deduce the following boundary condition for the
conformal factor $\psi$
\begin{equation}
  \label{eq:78}
  N_h \psi = \psi^3 \Theta_- + \psi^{-3} K_{ab} \nu^a \nu^b \text{ on }
  \partial \Omega,
\end{equation}
where $N_h \equiv 4\nu^aD_a  + H $. 

Equation \eqref{eq:78} is a boundary condition for the conformal
factor only. Remains to prescribe the boundary condition for the
momentum constraint (\ref{diver}). From \eqref{eq:78} we see that the
simpler way of solving our problem will be to chose for the momentum
constraint a boundary condition such that the function $K_{ab} \nu^a
\nu^b$ can be freely prescribed on $\partial \Omega$. Our first
result, concerning the momentum constraint, essentially says that this
is possible. In order to write this theorem we need to introduce some
additional concepts.  

The conformal Killing operator $\ck$, acting on vectors fields $w^a$,
is defined as
\begin{equation}
  \label{eq:4}
  (\mathcal{L}_h w)_{ab}=2D_{(a}w_{a)}-\frac{2}{3}h_{ab}D^cw_c.
\end{equation}
We say that $\xi^a$ is conformal Killing vector field if it satisfies
$\ck \xi =0 $ on $\bar \Omega$. Given a conformal Killing vector field
$\xi^a$ we define the conformal Killing data at the point $i$ by
\begin{equation}
\label{eq:ckvdata}
k_a = \frac{1}{6}\,D_a D_b \xi^b(i), \,\,
S^a={\epsilon^a}_{bc}D^b \xi^c(i), \,\,  
q^a=\xi^a(i),\,\, 
a =\frac{1}{3} D_a\xi^a(i). 
\end{equation}
Since $\Omega$ is connected, the integrability conditions for
conformal Killing fields (cf. \cite{Yano57}) entail that these ten
conformal Killing data at $i$ determine the field $\xi^a$ uniquely on
$\Omega$.

Because of the singular behavior of $K^{ab}$ at $i$ given by
(\ref{eq:77}), it will convenient to write the results in terms of a
further rescaled metric $h'_{ab}$ for which the curvature vanished at
$i$. The metric $h'_{ab}$ can be explicitly computed as follows.
Denote by $B_\epsilon$ the open ball with center $i$ and radius $r =
\epsilon > 0$, where $\epsilon$ is chosen small enough such that
$B_\epsilon$ is a convex normal neighborhood of $i$.  We define the
cut-off function $\chi_\epsilon$ as a non-negative, smooth function
such that $\chi_\epsilon = 1$ in $B_{\epsilon/2}$ and $\chi_\epsilon = 0$ in
$\Omega \setminus B_\epsilon$.  Consider the following smooth conformal
factor
\begin{equation}
\label{eq:cf}
\omega_0 = e^{\chi_\epsilon f_0} 
\text{ with } f_0 = \frac{1}{4}\,x^j x^k\,L_{jk}(i),
\end{equation}
where we have used the value at $i$ of the tensor $L_{ab}\equiv R_{ab}
-\frac{1}{4}R h_{ab}$.  By a straightforward calculation we get that
the Ricci tensor of the metric
\begin{equation}
 \label{eq:h00}
h'_{ab}=\omega_0^4 h_{ab}
\end{equation}
vanishes at the point $i$. 

Equation (\ref{diver}) is solved using the standard York splitting
(cf. \cite{York73}) adapted to our setting. The free data is a trace
free, symmetric, tensor $Q^{ab}$. Since we have two kinds of boundary
conditions, $Q^{ab}$ has a natural decomposition
\begin{equation}
  \label{eq:58}
 Q^{ab}= Q^{ab}_{sing} + Q^{ab}_{reg},
\end{equation}
where $\Qs{ab}$ and $\Qr{ab}$ can be roughly characterized as follows
(see section \ref{mc} for details). 

The tensor $\Qs{ab}$ is different
from zero only in $B_\epsilon$ and pick up the singular behavior of
$K^{ab}$ at $i$, i.e; it will blow up like $r^{-4}$ at $i$. It
contains the linear and angular momentum of the data. These physical
quantities appear as constants $P^a$ and $J^a$ which partially
characterize $\Qs{ab}$. In addition to $P^a$ and $J^a$, there are
other constants $Q^a$ and $A$ involved in $\Qs{ab}$; these ten
constants are related with the conformal Killing data
\eqref{eq:ckvdata} as we will see.  

The tensor $\Qr{ab}$ gives the
boundary value of $K^{ab}$ at $\partial \Omega$ and will not
contribute to the linear and angular momentum.  For example we can
take $\Qr{ab}$ smooth in $\bar \Omega$; however a more general
behavior at $i$ is allowed, in particular $\Qr{ab}$ can blow up like
$r^{-1}$ at $i$. 

With these definitions, we can write our existence
theorem for the momentum constraint.
\begin{theorem}\label{T1}
  Assume $h_{ab}$ satisfies (\ref{eq:h-reg}) and $\partial \Omega$ is
  smooth.  Let $Q^{ab}$ the symmetric, trace free tensor, given by
  (\ref{eq:58}) where $\Qr{ab}$ and $\Qs{ab}$ satisfy
  (\ref{eq:62})--~(\ref{eq:41}) and (\ref{eq:Qn}) respectively.
  
  i) If the metric $h_{ab}$ admits no conformal Killing fields on $\Omega$,
  then there exists a unique vector field $w^a \in C^\infty(\bar
  \Omega \setminus \{i \} )\cap \hk{1}$ such that the tensor
  $K^{ab}$ defined by 
\begin{equation}
    \label{eq:59}
    K^{ab} = \omega_0^{10}\,(Q^{ab} - (\mathcal{L}_{h'} w)^{ab}),
  \end{equation}
  satisfies the equation $D_a K^{ab}=0$ in $\tilde \Omega$ and
  $K^{ab}\nu_a=\Qr{ab}\nu_a$ on $\partial \Omega$.  The metric
  $h'_{ab}$ and the conformal factor $\omega_0$ are defined by
  (\ref{eq:h00}) and (\ref{eq:cf}).

ii) If the metric $h_{ab}$ admits conformal Killing fields $\xi^a$ on
$\Omega$, a vector field $w^a$ as specified above exists if and only
if the constants $P^a$, $J^a$, $A$ and $Q^a$ (partly) characterizing
the tensor field $Q_{sing}^{ab}$, satisfy the equation
\begin{equation}
\label{eq:sycond}
P^a\,k_a + J^a\,S_a + A\,a + (P^c\,L_c\,^a(i) + Q^a)\,q_a =
\int_{\partial \Omega} Q^{ab}_{reg}\xi_a \nu_b \, dS,    
\end{equation}
for any conformal Killing field $\xi^a$ of $h_{ab}$; where the constants
$k^a$, $S^a$, $a$ and  $q^a$ are the conformal Killing data
at $i$ for  $\xi^a$ given  by (\ref{eq:ckvdata}).\\    
In both cases i) and ii) $K^{ab}$ is unique,  $K^{ab}\in
C^\infty(\bar \Omega \setminus\{i \} )$ and  $K^{ab}= O(r^{-4})$ at $i$.  

\end{theorem}

Note that in case ii) $K^{ab}$ is unique but $w^a$ is not, we can add
to a given solution any conformal Killing vector. Condition
\eqref{eq:sycond}, which arise in the presence of conformal
symmetries, is the natural extension to the analog condition discussed in
\cite{Beig96} and \cite{Dain99} where there is no inner boundary and in
\cite{Dain02b} where there exists matter sources in a compact
region. The fact that not only Killing vectors but also conformal
Killing vectors appear in \eqref{eq:sycond} is a particular feature of
maximal slices (i.e; $\tilde K = K=0$), see the discussion in
\cite{Dain02f}.  

The next result is concerning the Hamiltonian constraint
\eqref{Lich}. In order to enunciate it, we need two auxiliary
solutions. The first one, denoted by $\psi_0$, is a
solution of the problem with $K^{ab}=0$ on $\Omega$ (i.e; time symmetry) and
$\Theta_+=\Theta_-=0$ on $\partial \Omega$. That  is, $\psi_0$ satisfies the
following linear boundary value problem
\begin{align}
  \label{eq:7}
  L_h\psi_0 &=0 \text{ in } \tilde \Omega,\\
  \label{eq:11}
  N_h\psi_0 &=0 \text{ on } \partial \Omega,\\
  \label{eq:12}
  \lim_{r\rightarrow 0} r\psi_0 &=1 \text{ at } i.
\end{align}
The second particular solution, $\psi_1$, is a solution of the
following linear boundary value problem 
\begin{align}
  \label{eq:7b}
  L_h\psi_1 &= -\frac{1}{8}K_{ab}K^{ab}\psi_0^{-7}\text{ in } \tilde \Omega,\\
  \label{eq:11b}
  N_h\psi_1 &= \psi_0^{-3} K_{ab} \nu^a \nu^b \text{ on } \partial \Omega,\\
  \label{eq:12b}
  \lim_{r\rightarrow 0} r\psi_1 &=1 \text{ at } i.
\end{align}

\begin{theorem}\label{T2}
  Assume that the conformal metric $h_{ab}$ satisfies \eqref{eq:h-reg}
  and that $\partial \Omega$ is smooth. Assume also that $R\geq 0$ in
  $\bar \Omega$, $H\geq 0$ on $\partial \Omega$ and that either $R$ or
  $H$ is not identically zero. Then:
  
  i) There exist a unique, positive, solution $\psi_0$ of
  \eqref{eq:7}--\eqref{eq:12}, and $\psi_0\in C^\infty(\bar \Omega
  \setminus \{i \} )$.
  
  ii) Let $K^{ab}$ be given by theorem \ref{T1}. Assume, in addition,
  that 
  \begin{equation}
    \label{eq:vv}
Q_{reg}^{ab} \nu_a \nu_b \geq 0 \text{ on } \partial \Omega.    
  \end{equation}
  Then there exist a unique, positive, solution $\psi_1$ of
  \eqref{eq:7b}--\eqref{eq:12b}, and $\psi_1\in C^\infty(\bar \Omega
  \setminus \{i \} )$.
  
  iii) Let $\Theta_-\leq 0$ and assume it satisfies the following
  inequality
  \begin{equation}
    \label{eq:80}
    |\Theta_-| \leq  \psi_1^{-6} Q_{reg}^{ab} \nu_a \nu_b.
  \end{equation}
  Then there exist a unique, positive, solution $\psi$ of equation
  \eqref{Lich} with boundary conditions \eqref{eq:17} and
  \eqref{eq:78}, and $\psi\in C^\infty(\bar \Omega \setminus \{i \}
  )$.  Moreover, $\psi_0\leq \psi \leq \psi_1$ and
  \begin{equation}
    \label{eq:76}
    \Theta_+\leq \Theta_-\leq 0.
  \end{equation}
\end{theorem}

Let us discuss the assumptions and the conclusions of this theorem. We
have assumed that the boundary $\partial \Omega$ is smooth, this is
done for simplicity and because there seems not be any physical reason
to study trapped surfaces with rough boundary. For rough boundaries
(for example with corners or cusps) the following proofs are not
valid.

The condition $R\geq 0$ and $D_a\nu^a \geq 0$ ensures that the
conformal metric $h_{ab}$ can be conformally rescaled to a metric
$\hat h_{ab}$ for which the boundary $\partial \Omega$ is an extremal
surface (i.e; $\hat H=0$) and $\hat R>0$. Then, this condition can
expressed in terms of the corresponding Yamabe class studied in
\cite{Escobar92}. A similar condition also appears in the case without
inner boundary, see \cite{Cantor81} and also \cite{Friedrich98},
\cite{Dain99} for a discussion of the same condition using a
compactification as we use here.

Simple examples of sets $(h,\Omega)$ which satisfy these conditions
are the followings. Take $h_{ab}=\delta_{ab}$, where $\delta_{ab}$ is
the flat metric. Let $\Omega=B_{r_0}$, where $B_{r_0}$ is a ball of
radius $r_0$. Then $H=2/r_0$ and $R=0$. This example has only one
boundary component. Note that, in this example, we can not take a
second ball inside $B_{r_0}$ to construct a region like the one showed
in Fig. \ref{fig:2}, because in this case on the boundary of the
interior ball $H$ will be negative, owing to our choice of the normal
$\nu^a$. To construct an example with two boundary components, take
$h_{ab}= \omega^4 \delta_{ab}$, where $\omega=1+1/r$; and let $\Omega$
be the annulus $\Omega=B_{r_0}-B_{r_1}$ where $r_0>1>r_1>0$. Then we
have that $R=0$, $H=2(1+1/r_0)^{-3}(1-1/r_0)/r_0>0$ on $\partial
B_{r_0}$ and $H=2(1+1/r_1)^{-3}(1/r_1-1)/r_1>0$ on $\partial B_{r_1}$.
Note that in this example the point $i$ can not be at the origin
$r=0$. The generalization to $k$ boundaries components is
straightforward.

Condition \eqref{eq:vv} can be written in terms of physical quantities
as $\tilde K^{ab}\tilde \nu_a \tilde \nu_b\geq 0$. From equations
\eqref{eq:21}--\eqref{eq:21b} we see that, remarkably, \emph{every
  future marginally trapped surface on a maximal slice satisfies this
  condition}. In contrast, condition \eqref{eq:80} is a sufficient
condition which is not expected to be also necessary in general.

A future trapped surface is a time asymmetric concept.  If we reverse
the time direction $t^a \rightarrow -t^a$  we have $l^a
\rightarrow -k^a$ and $k^a \rightarrow -l^a$ and hence $\Theta_+
\rightarrow - \Theta_-$, $\Theta_- \rightarrow - \Theta_+$. Then, if
we have a future trapped surface in one time direction, we will have
$\Theta_+, \Theta_- >0$ in the opposite time direction and then it
will be not trapped in this time direction. 
The singularity theorems \cite{Penrose65} implies that the development
of a data which contain a future trapped surface will be geodesically
incomplete into the future, but nothing is said about the past. In
fact, in a realistic gravitational collapse, the spacetime is expected
to have no singularities in the past. On the other hand, if the data
have non-trivial topology (as the ones usually used in numerical
simulations, see for example \cite{Cook00}) we can apply the
singularity theorem given in \cite{gannon75} to conclude that the
spacetime will be geodesically incomplete in both time directions.
This shows that the class of data constructed here is more general
that the one with non-trivial topology, since it  does not rule out
data with regular past.

It is perhaps surprising that $\Theta_-$ and not $\Theta_+$ is free
data at the boundary. For example, we can not prescribe $\Theta_+=0$
and $\Theta_-<0$ at the boundary, since this will contradict
inequality \eqref{eq:76}. If we start with a boundary that satisfies
$\Theta_+,\Theta_-<0$, then, in general, a surface with $\Theta_+=0$,
$\Theta_-<0$ will be present in the interior of $\tilde \Omega$, this
will correspond, for example, to the outer most trapped surface. But
we can not use theorem \ref{T2} to start with such a surface at the
boundary, the location of this surface can not be given a priori under
our assumptions. On the other hand, theorem \ref{T2} allows to
prescribe $\Theta_+=0$ at the boundary, this is just the time
inversion of the case $\Theta_-=0$, $\Theta_+\leq 0$ which gives
$\Theta_+=0$, $\Theta_-\geq 0$. However, unless $\Theta_-=0$, these
surfaces \emph{will not} be future marginally trapped. Moreover, these
surfaces are located on the inner null boundary of the left quadrant
of the Kruskal diagram (region IV of Fig. 6.9 in \cite{Wald84}), and
hence they are not expected to be present in any realistic
gravitational collapse of matter.

In the Kruskal diagram, trapped surfaces which satisfy condition
\eqref{eq:76} are located in the left half of region II. Hence, they
can be present in a spherically symmetric collapse. However, condition
\eqref{eq:76} is a restriction in the sense that not in every collapse
such surfaces will exist, as can be shown using the spherically
symmetric example.

The choice of $\Theta_-$ as a free data is dictated by the signs in
the boundary condition \eqref{eq:78}. The boundary condition should
have the appropriate signs to use the maximum principle in order to
prove that the conformal factor $\psi$ is positive (see sections
\ref{pre} and \ref{hc}).

Finally, we note that theorems \ref{T1} and \ref{T2} show how to
obtain the whole solution in the exterior region $\tilde \Omega$
solving the equations in the compact domain $\Omega$. This can be used
in numerical computations to solve for the whole data on a finite grid
(see for example \cite{Husa96}, \cite{Husa98b}).\footnote{After this
  work was completed, there has appeared an article by D. Maxwell
  \cite{Maxwell03} which studies the boundary condition $\Theta_+=0$.
  In a new version of this article, which appeared after this work was
  submitted for publication, Maxwell makes an important improvement
  and he is now able to construct solutions which satisfy
  $\Theta_-\leq \Theta_+=0$ at the boundary. This is done imposing an
  extra condition which involves $H$ and $K_{ab}\nu^a\nu^b$.
  Remarkably, the solutions obtained in \cite{Maxwell03} and the ones
  obtained in the present article do not, in general, overlap.
  However, it is important to emphasize that both set of solutions do
  not cover all possible black hole exterior regions.  It is still an
  open problem how to construct and characterize all initial data for
  black holes exterior regions.

I am grateful to D. Maxwell for useful discussions.}

\section{Preliminaries}\label{pre}
In this section we summarize some results from functional analysis and
the theory of linear elliptic partial differential equations which we
will use in the following proofs. Some of these results, although
standard, are not easily available in the literature.

The operators $L_h$ (defined after Eq. \eqref{Lich}) and $\Lh$
(defined by \eqref{eq:6}) which appear in the Hamiltonian and Momentum
constraints respectively, are linear, second order, elliptic operators
in divergence form.  We will use the Hilbert space approach to the
boundary value problem for this kind of operators. This approach  has
the advantage to be both simpler and applicable to a broader classes
of solutions than other methods.

Let $\Omega$ a bounded open domain in $\Rt$, it will be assumed in the
following that the boundary $\partial \Omega$ is \emph{smooth}.  We
shall use the following functions spaces (see \cite{Adams},
\cite{Gilbarg} for definitions, notations, and results) defined in
$\Omega$: the set of $m$ times continuously differentiable functions
$C^m(\Omega)$, the H\"older space $C^{m, \alpha}(\Omega)$, where
$0<\alpha<1$, the corresponding spaces $C^m(\bar \Omega)$, $C^{m,
  \alpha}(\bar \Omega)$, the space $C^\infty_0(\Omega)$ of smooth
function with compact support in $\Omega$, the Lebesgue space
$L^p(\Omega)$, the Sobolev space $\hk{s}$, and the local Sobolev space
$\hkl{s}$ where $s$ is a real number.

An elliptic operator in divergence form has an associated bilinear
form $\B$. In the particular cases of $L_h$ and $\Lh$ these bilinear
forms are given by \eqref{eq:38} and \eqref{eq:81}  respectively.  
A bilinear form is called \emph{weakly coercive} over $\hk{1}$ if
there exist  constants  $\lambda_1> 0$ and  $\lambda_0\geq  0$ such
that following inequality holds
\begin{equation}
  \label{eq:48}
  \B(v,v)\geq \lambda_1\hfk{v}{1}^2 - \lambda_0 \hfk{v}{0}^2
\end{equation}
for all $v\in \hk{1}$. If we can take  $\lambda_0=0$ in (\ref{eq:48})
we say that $\B$ is \emph{strictly coercive}. $\B$ is called
\emph{bounded}  in $\hk{1}$ if there exist a positive constant $C$
such that
\begin{equation}
  \label{eq:50b}
  |\B(v,u)| \leq C \hfk{v}{1} \hfk{u}{1}, 
\end{equation}
for all $u,v \in \hk{1}$. For a given $\B$, we define the following spaces
\begin{align}
  \spa{V} &= \{ u\in \spa{X} : \B(v,u)=0 \text{ for all } v\in  \hk{1}\},\\
  \spa{W} &= \{ u\in \spa{X} : \B(u,v)=0 \text{ for all } v\in  \hk{1}\}.
\end{align}
When $\B$ is symmetric we, of course, have $\spa{V}=\spa{W}$.

The following theorem is the basic existence tool for the linear
boundary problem, it is  a consequence of the Lax-Milgram theorem and
the Fredholm alternative in Hilbert spaces.
\begin{theorem} \label{exis}
  Let $\B$ be a bilinear form which is bounded and weakly coercive in
  $\hk{1}$. Then $\spa{V}$ and $\spa{W}$ have both finite dimension.
  Moreover, let $F$ be a bounded linear functional over $\hk{1}$.  Then,
  there exist a function $u\in \hk{1}$ such that
  \begin{equation}
    \label{eq:19}
    \B(v,u)=F(v) \text{ for all } v \in  \hk{1},
  \end{equation}
if and only if 
\begin{equation}
  \label{eq:30}
  F(w)=0,
\end{equation}
for all $w\in \spa{W}$.
\end{theorem}
We will apply this theorem for both second order elliptic equations
and second order elliptic systems. Although the methods in both cases
are, in many aspect,  similar, second order elliptic equations
have distinguished properties,   we will treat them separately.

Consider the operator $L_hu$.  If we multiply by $v\in C^1_0(\Omega)$
and integrate by parts we obtain the symmetric  bilinear form
\begin{equation}
  \label{eq:38}
  \B(u,v)=\int_\Omega \left ( h^{ab} D_au D_b v- R u v \right) d\mu,
\end{equation}
where $d\mu$ is the volume element with respect to $h_{ab}$. We have
the following result, which will allow us to use theorem \ref{exis}
for the existence proof.
\begin{theorem}\label{bc}
  Let $\B$ be given by (\ref{eq:38}). Assume that the metric satisfies
\eqref{eq:h-reg}. Then $\B$ is
  bounded and weakly coercive in $\hk{1}$.
\end{theorem}
For a proof see for example \cite{Gilbarg}, in \cite{ladyzhenskaya68}
and \cite{Chen98} the same result is proved under more general
assumptions on the coefficients.

Consider the following linear boundary value problem for the elliptic
operator $L_h$
\begin{align}
  \label{eq:61}
  L_hu &=f  \text{ in } \Omega , \\
\label{eq:61b}
 N_h u &= \varphi \text{ on }
\partial \Omega.  
\end{align}
Note that our assumption on $h_{ab}$ implies that $H\in
C^{\infty}(\partial \Omega)$.  A \emph{weak solution} of this problem
is a function $u\in \hk{1}$ which satisfies
\begin{equation}
  \label{eq:68}
  \B(u,v)+  \int_{\partial \Omega} H u v \, dS =-\int_\Omega
   fv \, d\mu +\int_{\partial \Omega} \varphi v\, dS,
\end{equation}
for all $v\in \hk{1}$, where $\B$ is given by \eqref{eq:38} and $dS$
is the surface element on $\partial \Omega$. One can
check that a smooth weak solution $u$ of \eqref{eq:68} will satisfies
\eqref{eq:61}--\eqref{eq:61b}. This kind of boundary conditions are
called natural boundary condition for the operator $L_h$, see
\cite{Agmon65}, \cite{ladyzhenskaya68} and \cite{Folland} for a
discussion on this method of treating general boundary problem.

We define the bilinear form $\B'$ by
\begin{equation}
  \label{eq:70}
  \B'(u,v)= \B(u,v)+  \int_{\partial \Omega}H u v \, dS,
\end{equation}
and the linear functional $F$ given by 
\begin{equation}
  \label{eq:71}
  F(v)=-\int_\Omega fv\, d\mu +\int_{\partial \Omega} \varphi v\, dS.
\end{equation}
That is, $\B'$ and $F$ are defined as the left and right hand side of
Eq. ~(\ref{eq:68}) respectively. We want to apply the abstract
existence theorem \ref{exis} for $\B'$ and $F$. This theorem is not
very useful unless we have a characterization of the null space
$\spa{W}$ of $\B'$. The maximum principle will be used to ensure that
$\spa{W}$ is trivial under additional assumptions on the coefficients.  

We have the following version of the maximum principle.
\begin{theorem}[Weak Maximum Principle]
\label{weakmaximum}
Let $\B'$ given by \eqref{eq:70}, where the metric satisfies
\eqref{eq:h-reg}. Assume that $R\geq 0$ and $H\geq 0$.     
Let $u\in \hk{1}$ satisfies $\B'(u,v) \leq (\geq  )0$ for all $v\in
\hk{1}$, $v\geq 0$. Then $u\leq  (\geq) 0$ in $\Omega$ or $u$ is a positive
(negative) constant. 
\end{theorem}
This theorem can be obtained from the more general results given in
\cite{trudinger77}. The proof is similar to  the proof of theorem 8.1 in
  \cite{Gilbarg}.  
  
  We will also use the following standard  versions of the maximum
  principle, see \cite{Gilbarg}.

\begin{theorem}[Strong Maximum Principle] \label{strongmaximum}
  Assume that the metric satisfies \eqref{eq:h-reg} and that $R\geq
  0$.  Let $u\in H^1(\Omega)$ satisfy $L_hu \geq 0$ in $\Omega$.  Then,
  if for some ball $B \subset \subset \Omega$ we have
\[
\sup_B u=\sup_\Omega u \geq 0,
\]
the function $u$ must be constant in $\Omega$.
\end{theorem}

\begin{theorem}[Hopf] \label{hopft}
Assume that  the metric satisfies
\eqref{eq:h-reg} and   that $R\geq 0$.
  Let $u\in C^2(\Omega)$. Suppose $L_h u \geq 0$ in $\Omega$.  Let
  $x_0\in \partial \Omega$ be such that
  \begin{itemize}
  \item[(i)] $u$ is continuous at $x_0$;
\item[(ii)] $u(x_0)> u(x)$ for all $x\in \Omega$ and $u(x_0)\geq 0$;
\item[(iii)] $\partial \Omega $ satisfies an interior sphere condition at
  $x_0$.
\end{itemize}
Then the outer normal derivative of $u$ at $x_0$, if it exist,
satisfies the strict inequality
\begin{equation}
  \label{eq:69}
  \frac{\partial u}{\partial \nu} >0.
\end{equation}
\end{theorem}

We can now prove the basic existence result for the elliptic linear
boundary value problem \eqref{eq:68}. 
\begin{theorem}\label{od}
  Assume that the metric satisfies \eqref{eq:h-reg}. Assume also that
  $R\geq 0$, $H\geq 0$ and that either $R$ or $H$ is not identically
  zero.  Let $f\in \lt$, $\varphi \in L^2(\partial \Omega)$. Then
  there exist a unique weak solution $u\in \hk{1}$ of the boundary
  value problem \eqref{eq:68}. If, in addition, we have $f\leq 0$,
  $\varphi\geq 0$, then $u\geq 0$.
\end{theorem}

\begin{proof}
  Let $\B'$ and $F$ be given by \eqref{eq:70} and \eqref{eq:71}. In
  order to apply theorem \ref{exis}, we need to prove that $F$ is
  bounded, $\B'$ coercive and bounded and finally that the
  corresponding null space $\spa{W}$ for $\B'$ is trivial. The
  boundedness of $F$ follows directly from Hölder inequality and the
  assumptions on $f$ and $\varphi$. 
  
  By theorem \ref{bc} we have that $\B$ is bounded, then to prove that
  also $\B'$ is bounded we only need to prove that the surface
  integral in (\ref{eq:70}) is bounded in $\hk{1}$. Using the
  generalized Hölder inequality we obtain
\begin{equation}
  \label{eq:72}
\left | \int_{\partial \Omega} H u v \, dS \right | \leq
||H||_{L^2(\partial
\Omega)}   ||u ||_{L^4(\partial
\Omega)}  ||v ||_{L^4(\partial
\Omega)}.     
\end{equation}
Since $u,v\in \hk{1}$, we can use the trace theorem (see \cite{Adams})
to conclude that $u,v\in H^{1/2}(\partial\Omega)$, by the imbedding
theorem  we have that $u,v\in L^4(\partial \Omega)$.
Then, we can replace the $L^4(\partial \Omega)$ norms in the left hand
side of (\ref{eq:72}) by $\hk{1}$ norms and hence the surface integral
is bounded in $\hk{1}$.

Using that $H \geq 0$ we have that $\B'(v,v)\geq \B(v,v)$, then the
coerciveness of $\B'$ follows directly from the coerciveness of $\B$
given by theorem \ref{bc}.

From the maximum principle \ref{weakmaximum} it follows that  
every  $w\in \spa{W}$ should be constant, by the assumption that
either $R$ or $H$ is not identically zero it follows that
$w=0$. The last statement about positivity of $u$ follows directly
from the maximum principle \ref{weakmaximum}.   
\end{proof}

We note that if we make the stronger assumption $H>0$ we don't
need to use the maximum principle in the previous proof, the
uniqueness followed directly in this case. 

The following global estimates for the non homogeneous case
will play a fundamental role in the non linear existence proof.
\begin{theorem}\label{w1p}
  Assume that the metric satisfies \eqref{eq:h-reg}.  Let $u\in
  \hk{1}$ be a weak solution of \eqref{eq:68}.  Assume also that $f\in
  \lt $ and $\varphi \in L^\infty(\partial \Omega)$. Then $u \in
  C^\alpha(\bar \Omega)$, $0< \alpha <1$.
\end{theorem}
The interior estimate part of this theorem (that is the fact that $u
\in C^\alpha(\Omega)$) is standard. The crucial part is the regularity
up to the boundary under the assumption $\varphi \in L^\infty(\partial
\Omega)$. This can be proved using Campanato spaces, following similar
arguments as in  \cite{Giaquinta93} and Theorem 3.8 of
\cite{han97}, where the interior estimate is proved.   

We turn now to the elliptic system defined by $\Lh w^a$. If we
multiply by a $u^a\in C^\infty_0(\Omega)$ we get the following
bilinear form defined over vectors
\begin{equation}
  \label{eq:81}
    \B(u,w)=\int_{\Omega}(\mathcal{L}_h w)^{ab} (\mathcal{L}_h u)_{ab} \, 
  d\mu.
\end{equation}
We have the following analog to theorem \ref{bc}.

\begin{theorem}\label{Garding}
 Assume that the metric satisfies \eqref{eq:h-reg}. Then $\B$
  defined by \eqref{eq:81} is bounded and weakly coercive on $\hk{1}$.
\end{theorem}
To prove this result, we first check that $\B$ satisfies the
Legendre-Hadamard conditions and then we use the G\aa rding inequality
for functions with compact support.  Using the assumption of the
smoothness of $\partial \Omega$ we can extend the result to $\hk{1}$
using the Sobolev extension theorems. See for example
\cite{Giaquinta93} for a general discussion of this kind of bilinear
forms.

Finally, for the non linear existence proof we will use the following
fixed point theorem (see for example \cite{Gilbarg}).

\begin{theorem}[Schauder fixed point] 
\label{Schauder}
 Let $B$ be a closed convex set in a Banach space $V$ and let $T$ be a 
continuous mapping of $B$ into itself such that the image $T(B)$ is
precompact, i.e. has compact closure in $B$. Then $T$ has a fixed point.
\end{theorem}

\section{The Momentum Constraint}\label{mc}
The aim of this section is to prove theorem \ref{T1}. The momentum
constraint is solved using the York splitting as follows.  Given a
trace free tensor $Q^{ab}$ we look for a vector field $w^a$ such that
the tensor $K^{ab}$ defined by
\begin{equation}
  \label{eq:3}
  K^{ab}= Q^{ab} -(\mathcal{L}_h w)^{ab}, 
\end{equation}
satisfies \eqref{diver}. Then, it follows that the vector $w^a$ will
satisfy the equation
\begin{equation}
  \label{eq:5}
  \mathbf{L}_hw_a=D^bQ_{ab} \text{ in } \tilde \Omega ,
\end{equation}
where 
\begin{equation}
  \label{eq:6}
  \mathbf{L}_hw_a=D^b(\mathcal{L}w)_{ab}.
\end{equation}

We want to prescribe $K^{ab}\nu_a \nu_b$ freely at the boundary
$\partial \Omega$, this suggests to impose the following boundary
condition for equation \eqref{eq:6} 
\begin{equation}
  \label{eq:2}
 (\mathcal{L}w)_{ab} \nu ^a=0  \text{ on } \partial \Omega,   
\end{equation}
where $\ck$ has been defined in \eqref{eq:4}. This boundary condition
satisfies the Lopatinski-Shapiro conditions required in the elliptic
estimates of \cite{Agmon64}.  Moreover, this is a natural boundary
condition for the operator $\Lh$ in the following sense. 
Assume that  $u^a$ and $w^a$ are sufficiently smooth vector fields, we
use 
\eqref{eq:6} and 
integrate by part to obtain the following identity
\begin{equation}
  \label{eq:34}
  \int_{\Omega} u_a \mathbf{L}_hw^a \, d\mu = \int_{\partial \Omega}
  (\mathcal{L}_h w)^{ab}\nu_a u_b \, dS -\B(u,w),  
\end{equation}
where  the bilinear form $\B(u,w)$ is given by \eqref{eq:81}. 
From equation \eqref{eq:34} we see that if $\Lh w^a=0$  in $\Omega$ and
$(\mathcal{L}w)_{ab} \nu ^a=0$  on  $\partial \Omega$, then $w^a$ is a
conformal Killing vector.  

For the moment let us not consider the singular behavior at $i$. Then
we have the following  boundary value problem
\begin{align}
  \label{eq:66}
\Lh w^a &= J^a \text{ in } \Omega , \\  
 (\ck w)_{ab} \nu ^a &= 0  \text{ on } \partial \Omega. \label{eq:66b}   
\end{align}
A vector field $w\in \hk{1}$ is a weak solution of  the boundary
value problem \eqref{eq:66}-\eqref{eq:66b} if 
\begin{equation}
  \label{eq:44}
\B(u,w)= - \int_{\Omega}u_a  J^a\, d\mu,  
\end{equation}
for all $u^a\in \hk{1}$.  Consistently with \eqref{eq:44}, we define a
conformal Killing vector field $\xi^a$  as weak solution of
\begin{equation}
  \label{eq:67}
  \B(u,\xi)=0,
\end{equation}
for all $u^a\in \hk{1}$.
We have the following existence theorem. 

\begin{theorem}
\label{m1}
Assume that the metric satisfies \eqref{eq:h-reg}.
  Let $J^a\in L^p(\Omega)$, $p\geq 6/5$ be a vector field
such that
\begin{equation}
    \label{eq:46}
    \int_{\Omega} J^a\xi_a \, d\mu=0, 
\end{equation}
for all conformal Killing vectors $\xi^a$ in $\Omega$.  
Then there exist a weak  solution $w^a \in \hk{1}$ of  
\eqref{eq:44}.
\end{theorem}
\begin{proof}
  We will use theorem \ref{exis}. By theorem \ref{Garding} we have
  that the bilinear form $\B$ defined by \eqref{eq:81}  is bounded and
  coercive in $\hk{1}$ under our assumptions on the metric $h_{ab}$.  Using
   Hölder inequality and the Sobolev imbedding theorem one can prove
  that the functional
  \begin{equation}
    \label{eq:39b}
    F(u)=\int_\Omega u_ a J^a \, d\mu,
  \end{equation}
  is bounded in $\hk{1}$ if $J^a \in L^p(\Omega)$, $p\geq 6/5$.
  Finally, condition (\ref{eq:46}) is just  equation $F(\xi)=0$ in
  theorem \ref{exis}.
\end{proof}
The boundary value problem \eqref{eq:66}-\eqref{eq:66b} has been
studied in elasticity, usually under stronger regularity assumptions
on the coefficients (see for example \cite{Marsden83}).

We analyze now the singular behavior at the point $i$.  As we point
out before, it will be convenient to take advantage of the conformal
invariance of the equations and to write them with respect to the
metric (\ref{eq:h00}).  Since the Ricci tensor of the metric
(\ref{eq:h00}) vanishes at $i$ and we are in three dimensions, the
Riemann tensor vanishes there too. Hence the connection and metric
coefficients of (\ref{eq:h00}) satisfy in normal coordinates ${x'}^k$
(with respect to $h'_{ab}$) centered at $i$
\begin{equation}
\label{eq:vR}
\Gamma'_i\,^j\,_k = O({r'}^2), \quad h'_{ij}=\delta_{ij} + O({r'}^3).
\end{equation}

In these coordinates let 
$\Psi^{ik}_{flat}\in C^\infty(B_\epsilon\setminus \{i\})$ 
be a trace free symmetric and divergence free tensor with respect
to the flat  metric $\delta_{kl}$, 
\begin{equation}
\label{eq:psi0}
\delta_{ik}\,\Psi^{ik}_{flat} = 0, \quad  
\partial_i\Psi_{flat}^{ik}=0 \text{ in } B_\epsilon\setminus \{ i\},
\end{equation}
with
\begin{equation}
\label{eq:o-4}
\Psi_{flat}^{ik} = O({r'}^{- 4}), \quad
\partial\Psi_{flat}^{ik} = O({r'}^{- 5}),\quad \partial
\partial\Psi_{flat}^{ik} = O({r'}^{- 6}). 
\end{equation}
All these tensors have been characterized in theorem 14 of
\cite{Dain99}. Denote by $Q^{ab}_{sing}$ the $h'$-trace free tensor
which is given  by
\begin{equation}
\label{eq:Qn}
Q^{ab}_{sing}= \chi_\epsilon \,(\Psi_{flat}^{ab}-\frac{1}{3}\,
h'_{cd}\,\Psi_{flat}^{cd}\,{h'}^{ab}).
\end{equation}
The tensor $\Qs{ab}$ has two important
properties. The first one is 
\begin{equation}
\label{eq:dviPhi-2}
D'_a Q^{ab}_{sing} = O({r'}^{-2}).
\end{equation}
This is a consequence of equations \eqref{eq:vR} and
\eqref{eq:psi0}. The second is the following. If the metric $h_{ab}$
admit a conformal Killing vector $\xi^a$, then 
\begin{equation}
  \label{eq:9}
  \int_{\Omega}\xi_b D_a \Qs{ab} \, d\mu = P^a\,k_a + J^a\,S_a + A\,a +
  (P^c\,L_c\,^a(i) + Q^a)\,q_a,  
\end{equation}
where the constants $k^a$, $S^a$, $a$ and $q^a$ are the conformal
Killing data at $i$ for $\xi^a$ given by (\ref{eq:ckvdata}) and the
constants $P^a$, $J^a$, $A$ and $Q^a$ partly characterize the tensor
field $Q_{sing}^{ab}$ (there exist two free functions that can be
freely prescribed in $Q_{sing}^{ab}$, see \cite{Dain99} for details).
Equation \eqref{eq:9} is written in terms of the metric $h_{ab}$, it
is possible to write it in term to the rescaled metric $h'_{ab}$, in
this case $L'_{ab}(i)=0$.

For the regular part  $Q^{ab}_{reg}$ we will assume  that
\begin{equation}
  \label{eq:62}
  Q^{ab}_{reg} \in C^{\infty}(\bar \Omega\setminus \{i \}),
\end{equation}
and
\begin{equation}
  \label{eq:41}
  Q^{ab}=O(r^{-1}), \quad \partial Q^{ab}=O(r^{-2}),
  \quad  \partial \partial Q^{ab}=O(r^{-3}).
\end{equation}
An example is of course $\Qr{ab}\in C^{\infty}(\bar \Omega)$, however
the most general condition \eqref{eq:41} matches naturally with the fall
off of $\Qs{ab}$ in the sense that the constant $Q^a$ gives the fall
of $O(r^{-2})$, the constants $A$ and $J^a$ the $O(r^{-3})$ one and the
constants $P^a$ the $O(r^{-4})$, see  \cite{Dain99}.

Using the fall off \eqref{eq:41} we obtain
\begin{align}
  \label{eq:14}
  \int_{\Omega}\xi_b D_a \Qr{ab} \, d\mu &= \int_{\partial \Omega}
  \Qr{ab}\xi_b \nu_a \, dS -\lim_{\epsilon \to 0 
    }\int_{\partial B_\epsilon} \Qr{ab}\xi_b n_a \, dS,\\
    &= \int_{\partial \Omega}
  \Qr{ab}\xi_b \nu_a \, dS.\label{eq:14b} 
\end{align}
We can now prove theorem \ref{T1}

\textbf{Proof of theorem \ref{T1}}

\begin{proof}
  Since the conformal factor $\omega_0$ defined by (\ref{eq:cf}) is
  smooth, we can write the equation  with respect to the
  metric $h'_{ab}$
\begin{align}
  \label{eq:66c}
  \mathbf{L}_{h'} {w'}^a &= J^a \text{ in } \Omega , \\
  (\mathcal{L}_{h'} w')_{ab} {\nu'}^a &= 0 \text{ on } \partial \Omega.
  \label{eq:66c2}
\end{align}
Define $J^a=D'_b Q^{ab}$. By the (\ref{eq:dviPhi-2}) and (\ref{eq:62})
we have that $J^a \in L^p(\Omega)$, with $6/5 \leq p<3/2$. Then we can
apply theorem \ref{m1}.  In the presence of conformal Killing vectors
the condition
$F(\xi)=0$ in theorem \ref{m1} can  be written as the equation
(\ref{eq:sycond}) using equations  \eqref{eq:9}  and \eqref{eq:14b}. This shows the
existence of $w^a \in \hk{1}$.

The regularity (i.e. $w^a \in C^\infty(\bar \Omega \setminus \{i \} )
$ and consequently $K^{ab} \in C^\infty(\bar \Omega \setminus \{i \} )
$ ) follows directly from the standard elliptic regularity theorems of
\cite{Agmon64}.
   
Remains to show that $K^{ab}$ satisfies 
\begin{equation}
  \label{eq:60}
  |K^{ab}|=O(r^{-4}).
\end{equation}
We have  that
\begin{equation}
  \label{eq:63}
  |K^{ab}|\leq |Q^{ab}| + |(\mathcal{L}_{h'} w)^{ab}) |.
\end{equation}
The first term on the right hand side of this inequality satisfies, by 
~(\ref{eq:o-4}) and \eqref{eq:41}, the estimate (\ref{eq:60}). We only
need to estimate the second term. Let $\Omega' \subset \subset
\Omega$, we have the following inequality
\begin{equation}
  \label{eq:64}
\sup_{x\in \bar \Omega'\setminus
  B_\epsilon} |(\mathcal{L}_{h'} w)^{ab})| \leq C ||w ||_{C^1(\bar
  \Omega'\setminus 
  B_\epsilon )}\leq C ||w ||_{H^{3}(\Omega'\setminus
  B_\epsilon )},    
\end{equation}
where the first inequality is obvious and the second is a consequence
of the Sobolev imbedding theorem. Using the interior  elliptic regularity
theorem of \cite{Agmon64} we have
\begin{equation}
  \label{eq:65}
  ||w||_{H^{3}(\Omega'\setminus
  B_\epsilon )} \leq C (||J||_{H^{1}(\Omega'\setminus
  B_\epsilon )}+||w||_{L^{2}(\Omega'\setminus
  B_\epsilon )}).
\end{equation}
Since $w\in \hk{1}$ then $||w||_{L^{2}(\Omega'\setminus
  B_\epsilon )} $ is clearly bounded for all $\epsilon$. For the first
term in the right hand side of \eqref{eq:65} we 
  use the assumptions  \eqref{eq:o-4}
and \eqref{eq:41} to conclude that 
\begin{equation}
  \label{eq:16}
  ||J||_{H^{1}(\Omega'\setminus
  B_\epsilon )}\leq \frac{C}{r^3},
\end{equation}
for every $\epsilon$. Combining the previous inequalities it follows
that
\begin{equation}
  \label{eq:18}
  \sup_{x\in \bar \Omega'\setminus
  B_\epsilon} |(\mathcal{L}_{h'} w)^{ab})| \leq C /r^3,
\end{equation}
for every $\epsilon$. Then, using \eqref{eq:63} the desired result
follows.
\end{proof}

\section{The Hamiltonian Constraint}\label{hc}
In this section we will prove theorem \ref{T2}. We begin with the item
i) of this theorem, that is the existence of the function
$\psi_0$ defined as a solution of the boundary value problem
\eqref{eq:7}--\eqref{eq:12}. This function will play an important role
in the non linear existence proof; it is essentially the Green
function of the operator $L_h$ with the boundary condition
\eqref{eq:11}.

In the following
theorem $\delta_i$ will denote the Dirac delta distribution with
source at $i$.

\begin{theorem}\label{green}
  Assume that the conformal metric $h_{ab}$ satisfies \eqref{eq:h-reg}
  and that $\partial \Omega$ is smooth. Assume also that $R\geq 0$ in
  $\bar \Omega$, $H\geq 0$ on $\partial \Omega$ and that either $R$ or
  $H$ is not identically zero.  Then, there exist a unique solution
  $\psi_0$ of the boundary value problem \eqref{eq:7}--\eqref{eq:12}.
  Moreover, $\psi_0$ satisfies
\begin{equation}
  \label{eq:13}
  L_h \psi_0=-4\pi \delta_i,
\end{equation}
in $\Omega$, $\psi_0>0$ in $\bar \Omega \setminus \{i\}$, $\psi_0\in
C^\infty( \bar\Omega \setminus \{i\}) $ and it has the following form
$\psi_0 = \chi_\epsilon/r + u$ with $u\in \hk{2}$.
\end{theorem}

\begin{proof}
Observing that $1/r$ defines a fundamental solution to the
flat Laplacian, we obtain  
\begin{equation}
\label{eq:Lchi}
  \Delta(\frac{\chi_\epsilon}{r})=-4\pi \delta_i+\hat \chi,
\end{equation}
where $\hat \chi$ is a smooth function on $\bar \Omega$ with support
in $B_\epsilon \setminus B_{\epsilon/2}$. The ansatz $\psi_0=\chi_\epsilon/r
+u$ translates the original equations into the following equations for $u$
\begin{align}
  \label{eq:8}
 L_h u &= -\hat L_h(\frac{\chi_\epsilon}{r}) - \hat \chi \text{ in } \Omega,\\
N_h u &= 0 \text{ on }\partial \Omega, \label{eq:8b}
\end{align}
where we have defined $\hat L_h$ by the expansion $L_h=\Delta+\hat
L_h$ in normal coordinates centered at $i$. A direct calculation shows
that $\hat L_h(\chi_\epsilon r^{-1}) \in \lt\cap C^\infty(\bar \Omega
\setminus \{i\})$. Then, by theorem \ref{od}, there exists a unique
solution $u\in \hk{1}$ of \eqref{eq:8}--\eqref{eq:8b}.  We can use the
standard elliptic regularity theorems to conclude that $u\in \hk{2}$,
which in particular implies (by the Sobolev imbedding theorem) that
$u\in C^0(\bar \Omega)$. We can further use the elliptic regularity in
the region $\bar\Omega \setminus \{i\}$ to conclude that $u \in
C^\infty( \bar\Omega \setminus \{i\})$.

To show that $\psi_0$ is strictly positive, we observe that it is
positive near $i$ (because $r^{-1}$ is positive and $u$ is bounded).
Take $\epsilon$ small enough such that $\psi_0$ is positive on
$\partial B_\epsilon$. Note that $L_h\psi_0=0$ and $\psi_0$ is smooth
in $\bar\Omega\setminus B_\epsilon$. Let $x_0\in\bar\Omega\setminus
B_\epsilon$ the point where $\psi_0$ reach its minimum in
$\bar\Omega\setminus B_\epsilon$.  Since $L_h\psi_0=0$, we can use the
strong maximum principle \ref{strongmaximum} to conclude that either
$\psi_0$ is constant in $\bar\Omega\setminus B_\epsilon$ or $x_0$ is
in the boundary of $\bar\Omega\setminus B_\epsilon$ and
$\psi(x)>\psi(x_0)$ for all $x\in \Omega\setminus B_\epsilon$. Since
we have assumed that either $R$ or $H$ is not identically zero,
$\psi_0$ can not be constant.
Assume  that $\psi_0(x_0) \leq 0$. Since $\psi_0>0$ on
$\partial B_\epsilon$, then it follows that $x_0\in \partial \Omega$.
Consider the function $-\psi_0$.
We will apply theorem \ref{hopft} on $\bar\Omega\setminus B_\epsilon$
for this function to get a contradiction. Hypothesis i), ii) and  iii) of
this theorem are satisfied. Using the boundary condition \eqref{eq:8b}
we have 
\begin{equation}
  \label{eq:73}
  4v^aD_a(-\psi)(x_0)=-H(-\psi)(x_0)\leq 0,
\end{equation}
which contradict \eqref{eq:69}. 
\end{proof}

We note that theorem \ref{hopft} is essential to prove the strict
inequality $\psi_0>0$, if  we use theorem \ref{weakmaximum} we only
get $\psi_0 \geq 0$.

We treat now the general, non linear, case. Set $\psi=\psi_0+u$, then
the boundary value problem given by \eqref{Lich}, \eqref{eq:17} and
\eqref{eq:78} can be written as follows
\begin{align}
  \label{eq:24}
  L_h u &= f(x,u) \quad  \text{ in } \Omega,\\
\label{eq:24b}
  N_hu &= \varphi(x,u) \quad \text{ on } \partial \Omega,
\end{align}
where
\begin{equation}
  \label{eq:29}
  f(x,u)=-\frac{K^{ab}K_{ab}}{8(\psi_0+u)^7},
\end{equation}
and 
\begin{equation}
  \label{eq:25}
  \varphi(x,u)=-(\psi_0+u)^3\varphi_1+ \frac{\varphi_2}{(\psi_0+u)^3},
\end{equation}
where we have used the notation
\begin{equation}
  \label{eq:43}
  \Theta_-=-\varphi_1, \quad K_{ab}\nu^a\nu^b=\varphi_2,
\end{equation}
to emphasize that the functions $\varphi_1$ and $\varphi_2$ in the
following are arbitrary, non-negative functions which are not
necessarily given by by \eqref{eq:43}.

We first recall some special properties of the functions $f(x,u)$
and $\varphi(x,u)$. The function $f(x,u)$ can be written as  
\begin{equation}
  \label{eq:30b}
  f(x,u_1)-f(x,u_2)=(u_1-u_2)\hat f(x,u_1,u_2),
\end{equation}
where 
\begin{equation}
  \label{eq:31}
  \hat f(x,u_1,u_2)= \frac{K^{ab}K_{ab}}{8}\sum_{j=0}^6
  (\psi_0+u_1)^{j-7}(\psi_0+u_2)^{-1-j}.
\end{equation}
Clearly $\hat f(x,u_1,u_2)\geq 0$ for any $u_1,u_2\geq 0$.
Also, for any $u\geq 0$ we have
\begin{equation}
  \label{eq:32}
  0\geq f(x,u)\geq f(x,0).
\end{equation}
For $\varphi(x,u)$ we have a similar expression 
\begin{equation}
  \label{eq:33}
  \varphi(x,u_1)-\varphi(x,u_2)=(u_2-u_1)\hat \varphi(x,u_1,u_2);
\end{equation}
where
\begin{equation}
  \label{eq:37}
  \hat \varphi(x,u_1,u_2)=  \sum_{j=0}^2 \varphi_1 
  (\psi_0+u_1)^{2-j}(\psi_0+u_2)^{j}+\varphi_2
  (\psi_0+u_1)^{-j-1}(\psi_0+u_2)^{-3-j}.
\end{equation}
If we assume  $\varphi_1,\varphi_2 \geq 0$,
then we have   that $\hat\varphi(x,u_1,u_2)\geq 0$ for any $u_1,u_2\geq 0$.
However, in contrast to $f(x,0)$, the function $\varphi(x,0)$
has, in principle, no definite sign.  

We define $u_+$ as a solution of the following linear problem
 \begin{align}
    \label{eq:45}
    L_h u_+ &=f(x,0) \text{ in } \Omega, \\
    N_h u_+ & =\varphi_2 \psi^{-3}_0  \text{ on } \partial
    \Omega.\label{eq:45c} 
  \end{align}
  If $\varphi_2\geq 0$ and $f(x,0)\in \lt$, then, by theorem \ref{od},
  there exist a unique solution $u_+\geq 0$ of
  \eqref{eq:45}--\eqref{eq:45c}. Note that the function $\psi_1$
  defined by \eqref{eq:7b}--\eqref{eq:12b} is given by
  $\psi_1=\psi_0+u_+$. The following theorem will prove items ii) and
  iii) of theorem \ref{T2}.

\begin{theorem}\label{h1}
  Assume that the hypothesis of theorem \ref{green} are in force. In
  addition, we assume that $\varphi_1,\varphi_2$ are smooth function
  on $\partial \Omega$ which satisfy $\varphi_1,\varphi_2 \geq 0$,
  $K_{ab}K^{ab}\psi_0^{-7} \in L^2(\Omega)$ and $\varphi(x,u_+)\geq 0$
  where $u_+$ is the unique solution of \eqref{eq:45}--\eqref{eq:45c}.
  Then, there exist a unique positive solution $u\in \hk{2}\cap
  C^\infty(\bar \Omega \setminus \{i \} )$ of the boundary value
  problem \eqref{eq:24}--\eqref{eq:24b}. Moreover, $0\leq u \leq u_+$.
\end{theorem}
\begin{proof}
  The proof is based on the Schauder fixed point theorem
  \ref{Schauder}.  
Define $B\subset C^0(\bar \Omega)$ as 
\begin{equation}
  \label{eq:47}
  B= \{u\in C^0(\bar \Omega): \, 0\leq u \leq u_+    \}. 
\end{equation}
Clearly $B$ is closed and convex subset of the Banach space $C^0(\bar
\Omega)$.  Let $w\in B$, we define the map $T(w)=u$ as follows. Let
$u$ be the unique solution given by theorem \ref{od}, of the following
linear boundary value problem
\begin{align}
    \label{eq:45b}
    L_h u &=f(x,w) \text{ in } \Omega \\
N_h u & = \varphi(x,w) \text{ on } \partial \Omega. 
  \end{align}
  Using equation \eqref{eq:33}, \eqref{eq:30b}, the maximum principle
  \ref{weakmaximum} and the condition $\varphi(x,u_+)\geq 0$ one shows
  that $T(B)\subset B$. Using the elliptic estimate \ref{w1p} we have
  that $u \in C^\alpha(\bar \Omega) $, and hence the image of $T(B)$
  is precompact because $C^\alpha(\bar \Omega)$ is compactly imbedded
  in $C^0(\bar \Omega)$. It is also clear that $T$ is continuous.
  Then, by the Schauder fixed point theorem \ref{Schauder}, there
  exist a fixed point $T(u)=u$ with $u\in C^0(\bar \Omega)$. The
  standard elliptic regularity implies that in fact $u\in \hk{2}\cap
  C^\infty(\bar \Omega \setminus \{i \} )$.  This finishes the
  existence part.

To prove uniqueness, let assume that there exist two solutions $u_1$
and $u_2$, the difference $u_1-u_2$ will satisfy the equations
\begin{align}
  \label{eq:20}
  L_h(u_1-u_2) &= (u_1-u_2)\hat f(x,u_1,u_2)\\
N_h(u_1-u_2) &=(u_2-u_1) \hat \varphi(x,u_1,u_2) \label{eq:20b}
\end{align}
where $\hat f$ and $\hat \varphi$ are given by \eqref{eq:31} and
\eqref{eq:37} respectively. Assume that $u_1\neq u_2$ in some set.
Since $u_1$ and $u_2$ are continuous functions, we can find a set
$\Omega'\subset \bar \Omega$ such that $u_1>u_2$ in $\Omega'$ (we
change $u_1$ by $u_2$ if necessary), $u_1-u_2=0$ on $\partial
\Omega'-\Gamma$, where $\Gamma$ denotes the portion of $\partial
\Omega'$ contained in $\partial \Omega$ and $u_1-u_2\geq 0$ on
$\Gamma$.  By \eqref{eq:20} we have $L_h(u_1-u_2)\geq 0$.  $u_1-u_2$
can not be a positive constant in $\Omega'$ because $\hat f\geq 0$ and
$R\geq0$. Then by the strong maximum principle there exist $x_0\in
\partial \Omega'$ such that $(u_1-u_2)(x_0)>(u_1-u_2)(x) $ for all
$x\in \Omega'$. The point $x_0$ can not be on $\partial
\Omega'-\Gamma$ because there we have $u_1-u_2=0$ and this will
contradict $u_1>u_2$. Then $x_0\in \Gamma$. Using the boundary
condition \eqref{eq:20b} we get
\begin{equation}
  \label{eq:74}
  4\nu^aD_a(u_1-u_2)(x_0)\leq -H (u_1-u_2)(x_0)\leq 0.
\end{equation}
 We use theorem \ref{hopft} to get a contradiction.  
\end{proof}
Note that $\varphi(x,u_+)\geq 0$ is precisely condition
\eqref{eq:80}. To finish the proof of the item iii) of theorem
\ref{T2} it remains only to prove the inequality \eqref{eq:76}. 
 Using equations
\eqref{eq:t+}--\eqref{eq:t-} we obtain
\begin{equation}
  \label{eq:53}
  \Theta_- - \Theta_+=2N_h\psi=2\varphi(x,u)\geq 2\varphi(x,u_+)\geq 0.
\end{equation}

Using the sub and super solution method, boundary value problem of the
form \eqref{eq:20}--\eqref{eq:20b} has been studied in \cite{Amann71}
under stronger regularity assumptions on the coefficients (see also
\cite{pao92}). In \cite{herzlich97} a related equation has been also
studied using a variational approach.

\section*{Acknowledgments}
It is a pleasure to thank H. Friedrich, M. Mars and B. Schmidt for
illuminating discussions. I would also like to thank the organizers of
the ``Penrose inequalities workshop'', R. Beig, P. Chrusciel and W.
Simon, and the friendly hospitality of the Erwin Schrödinger Institute
for Mathematical Physics (ESI), where part of this work was done.

This work has been supported by the Sonderforschungsbereich SFB/TR7
of the Deutsche Forschungsgemeinschaft.


\end{document}